\documentstyle[budapest1]{article}  
\def\rightmark{The reconstructed final state of Au+Au collisions at RHIC-1} 
\def\leftmark{T. Cs\"org\H{o} and A. Ster }

\def\bea{\begin{eqnarray}}
\def\eea{\end{eqnarray}}
\def\be{\begin{equation}}
\def\ee{\end{equation}}

\def\dst{\displaystyle\phantom{|}}
\def\ov{\over\displaystyle}

\def\l({\left(}
\def\r){\right)}
\def\ave#1{\langle #1 \rangle}

\title{\bf The reconstructed final state of $Au+Au$ collisions\\
	from PHENIX and STAR data at $\sqrt{s} = 130 $ AGeV\\
	- indication for quark deconfinement at RHIC
} 
\authors{
{T. Cs\"org\H{o}$^{1}$ and A. Ster$^{1,2}$
}\\[2.812mm]
{\normalsize
\hspace*{-8pt}$^1$ MTA KFKI RMKI, Konkoly Thege 29-33,\\ 
H - 1121 Budapest, Hungary\\[0.2ex] 
\hspace*{-8pt}$^2$ MTA KFKI MFA, Konkoly Thege 29-33,\\ 
H - 1121 Budapest, Hungary
}}
 
\abstract{ 
        The final state of $Au + Au$ collisions 
	at $\sqrt{s}=130$ AGeV at RHIC has been 
        reconstructed within the framework of the Buda-Lund hydrodynamical
	model, by performing a simultaneous fit to {\it final} 
        data on two-particle Bose-Einstein correlations
	of the STAR and PHENIX Collaborations, and {\it final} identified
	single particle spectra as measured by the PHENIX Collaboration. 
	The  results indicate a strongly three dimensional expansion, with a 
	four-velocity field that is almost a
	spherically symmetric  Hubble flow. 
	We find large transverse geometrical source sizes,
	$R_G = 9.8 \pm 1.2$ fm, relatively short mean freeze-out time,
	$\tau_0 = 6.1 \pm 0.3$ fm/c and a
	short duration of particle emission, $\Delta \tau = 0.02 \pm 1.5$
	fm/c.  Most strikingly, we find an indication for a hot central part 
	of the hydrodynamically evolving  core,  
	characterized by a central temperature of
	$T_0 = 202 \pm 13 $ MeV, that is close to (or even  above)
	the deconfinement temperature of the quark-hadron
	phase transition.  The best fit indicates a cold surface 
	temperature of $T_s = 110 \pm 16$ MeV.  When the possibility of 
	the hot center is excluded, the confidence level of the 
	fit decreases from 28.9\% to 1.0 \%.
	Predictions are made for the rapidity dependence of 
	the slope parameters and for the transverse mass dependence of the 
	rapidity width of the single particle spectra, 
	and the transverse velocity dependence of the non-identical
	particle correlations.
  }

 
\begin{document}
 
\maketitle
%\setcounter{page}{1}
%\vfill\eject
\section{Introduction}\label{intro}
	The reconstruction of hadronic final state from the
	measured single particle spectra and two-particle 
	correlation functions is of great current research 
	interest in high energy
	heavy ion collisions. The goal of these studies is
	to characterize the final state with the help of a few 
	simple parameters like the mean freeze-out temperature
	or the mean transverse flow, or the geometrical size of the
	system at the mean freeze-out time. From the properties
	of the hadronic final states, one aims to identify  
	one (or more) new phases of hot and dense hadronic matter
	in the collisions of the biggest nuclei at the highest 
	available bombarding energies. 
	
	Space-time picture reconstruction of particle emitting sources
	in high energy physics is based on improved methods of
	intensity interferometry,
	a technique invented originally  by the radio astronomers
	R. Hanbury Brown and R. Q. Twiss~\cite{hbt}
	 to measure the angular diameter
	of main sequence stars. Intensity correlations
	in high energy physics are measured in momentum space,
	in contrast to stellar intensity interferometry, where the
	correlations are determined in the coordinate space. 
	Intensity correlations of identical
	particles appear due to quantum statistics as well as Coulomb
	and strong final state interactions. Dominated by quantum statistical
	effects, correlations of identical bosons are frequently referred
	to as Bose-Einstein correlations (and the name of fermionic
	correlations is Fermi-Dirac correlations). 
	The radius parameters of the two-particle Bose-Einstein correlation
	functions are frequently referred to as HBT radii to honor
	Hanbury Brown and Twiss.  In particle physics, 
	correlations of pions with small opening angles
	were observed first by G. Goldhaber, S. Goldhaber, A. Pais and W. Lee,
	~\cite{gglp} hence they are also referred to as GGLP correlations.
	
	Our tool for the space-time picture reconstruction
	is the Buda-Lund hydro (BL-H) model, introduced in refs.~\cite{3d,3dqm}
	and reviewed recently in ref.~\cite{cs-review}. 
	See also refs.~\cite{weiner-review,uli-review,bl-datong} for recent
	reviews on particle interferometry in high energy physics.

	First of all, let us remind the readers about 
	the difficulty of the interpretation of the experimental 
	data on correlations and spectra in
	high energy heavy ion collisions.
	High expectations seem to exist each time, when a new accelerator 
	starts its data taking. Our hopes suggest that we enter a whole
	new land to explore, and we tend to forget about the landscape
	that we have just left behind. 

	Let us remember, that the ``RHIC HBT puzzle"~\cite{gyulassy-puzzle} 
	is in fact not a RHIC specific phenomena:
	a similar discrepancy between predictions and final data 
	prevailed already at CERN SPS, 
	see refs.~\cite{ferenc-puzzle,na44-kk-prl}. 
	Essentially, the RHIC HBT puzzle is that the ratio of the
	outward and the sideward HBT radii was measured to be 
	$R_{\rm o}/R_{\rm s} \approx 1.0$, see refs.
	~\cite{gyulassy-puzzle,star_hbt,phenix_hbt}, and this result 
	was in contrast to certain theoretical expectations. 
	The ``CERN SPS HBT puzzle" could be similarly formulated:
	Why the difference between the outward and the sideward
	HBT radii is zero at CERN SPS? 
	This question can be based on Fig. 4 of ref.~\cite{ferenc-puzzle},
	see ref.~\cite{na44-kk-prl} for the most recent
	data with such a behavior in Pb+Pb collisions at CERN SPS,
	valid not only for pions but also for kaons.
	At CERN SPS, the situation seems  to be strikingly similar to 
	the happenings at RHIC. Perhaps the story at CERN SPS
	is so old by now, that its moral has been almost forgotten.

	Apparently, failed predictions of the RHIC HBT data can be assigned to
	models that were not tuned to  successfully
	describe the single particle spectra and the two-particle
	Bose-Einstein correlation functions at CERN SPS.
	On the other hand, models that worked well at CERN SPS 
	describe RHIC spectra and HBT data rather well, see e.g. 
	refs.~\cite{humanic-sps,humanic-rhic,na22,ster-qm99,ster-jhep,fb,fb-2}.

	There were two classes of predictions for the measurable
	HBT radius parameters in $Au + Au$ collisions at RHIC. 
	The predominant expectation was, that a soft, long lived,
	evaporative quark-gluon plasma phase will be produced,
	and the large duration of the particle emission can be
	observed from the big increase of the out component
	of the Bose-Einstein correlation functions, $R_{\rm out} 
	\gg R_{\rm side}$, a phenomena first observed
	by S. Pratt in refs.~\cite{pratt-qgp1,pratt-qgp2}.
	Similarly motivated calculations, with more realistic
	geometry and initial conditions were performed by
	Bertsch~\cite{bertsch-qgp1} and collaborators as well as by 
	Gyulassy and Rischke~\cite{rischke-plot}.
 
	However, there was another, less well known
	class of predictions for RHIC: 
	instead of predicting  
	$R_{\rm out} / R_{\rm side} \gg 1$, 
	refs.~\cite{csorgo-csernai,3d,3dqm} predicted 
	$R_{\rm out} / R_{\rm side} \approx 1$ . 
	It is also mentioned there
	how the sudden freeze-out of hadrons is related to
	an explosive particle production from a supercooled quark-gluon
	plasma~\cite{csorgo-csernai}, or a quark matter~\cite{csorgo-qm}, where 
	the gluonic degrees of freedom are not active. 
	Such a picture may emerge from a quasi-particle picture
	of a QGP, where the quarks and the gluons dress up in the
	vicinity of the phase transition, the constituent
	quark mass being of the order of 300 MeV, with a
	gluon mass $m_g \gg T_c$,~\cite{levai-heinz}. The sudden,
	explosive particle production, the hard equation of state
	and the quark combinatorics of particle yields~\cite{alcor}
	supports such a scenario at RHIC. 

	 As the characteristic nucleation times of hadronic bubbles 
	inside a quark gluon plasma are of the order of 100 fm/c,
	which is an order of magnitude larger than the characteristic
	life-time of the expanding system, bubble formation is not
	fast enough to keep the system close to the Maxwell construction
	and near-equilibrium phase transition. Instead, a negative
	pressure state developes very soon which then decays due
	to its mechanical instability, the cavitation. Such process
	may happen through a time-like deflagration
	and a process was predicted
	to end in a {\it pion flash}, with a short, 1-3 fm/c duration
	of particle emission. See ref.~\cite{csorgo-csernai}
	for greater details and signatures of this process.

	In this manuscript, we present an indication 
	for the existence of a transient
	quark matter state at RHIC1.

	The structure of the body of the paper is as follows.
	In section \ref{s:2}, we highlight the most important
	fitting formulas of the BL-hydro model.
	In section \ref{s:3}, the fits to the single particle spectra 
	and the two-particle Bose-Einstein correlation functions are shown. 
	In section \ref{s:4} we summarize the results, 
	and discuss their interpretation
	of the results and the correlations 
	between the various fitting
	parameters. In section \ref{s:5} we predict the 
	rapidity dependence of the observables. 
	We also make a prediction for the transverse mass, and rapidity 
	dependence of non-identical particle correlation functions.
	Finally, we conclude.

\section{Single particle spectra and two particle correlations from
	the Buda-Lund hydro model\label{s:2}}

	Our direct aim is to reconstruct the hadronic final
	state from the measurable single-particle spectra and
	two-particle correlation functions.
	From this reconstructed final state
	and the knowledge of the equation of state of hot and dense hadronic 
	matter (e.g. from lattice QCD calculations) one can,
	in principle, reconstruct the initial state 
	of the reaction by running the 
	(relativistic) hydrodynamical equations backwards in time, and
	determine if this initial state had been in the 
	QGP phase or not. Here we report on such 
	a direct reconstruction of the hadronic final state
	within the framework of the Buda-Lund hydro (BL-H) model,
	but we do not consider the more indirect reconstruction
	of the initial state.

	The BL-H model is a hydrodynamical parameterization
	of the hadronic final state, that has to be clearly distinguished
	from  a fully developed, time dependent solution of relativistic
	hydrodynamics. However, the BL inspired a whole new series
	of non-relativistic and relativistic, simple analytic solutions
	of fireball hydrodynamics. The non-relativistic hydro solutions
	correspond to the non-relativistic limit of the BL hydro model,
	with time-dependent model parameters,~
	\cite{cspeter,rsol,ellsol,ellobs,cssol,infl}.
	It is very interesting to note, that the governing equations for
	the scale parameters are similar to that of the 
	Zim\'anyi-Bondorf-Garpman hydrodynamical solution~\cite{zbg}
	and its ellipsoidally symmetric generalization~\cite{dgsbz}.
	
	However, in the relativistic domain, 
	only the coasting (accelerationless) 
	hydro solutions were found analytically 
	until now~\cite{rel-1d,rel-cyl,rellsol}, in an attempt
	to figure out the governing equations for the BL-H model parameters. 
	Although Bjorken's well known solution of relativistic hydrodynamics 
	~\cite{bjorken}
	belongs also to this class of accelerationless solutions, 
	and the new coasting relativistic solutions
	include finite 1+1 dimensional solutions~\cite{rel-1d}, 
	1+3 dimensional
	solutions with cylindrical~\cite{rel-cyl} and ellipsoidal symmetry
	~\cite{rellsol}, the search is still going on for even more realistic, 
	relativistic hydro solutions that can interpolate
	from the non-relativistic domain to the relativistic one 
	even if the acceleration of the matter is significant. 

	Furthermore, the Buda-Lund flow profile, with a
	time-dependent radius parameter $R_G$, 
	was recently shown to be an exact solution of relativistic 
	hydrodynamics of a
	perfect fluid at a vanishing speed of sound~\cite{biro}.   
	It turned out~\cite{ellobs,cssol}, 
	that the flow field is a generalized Hubble flow and
	the average transverse flow at the geometrical radius is 
	formally similar to Hubble's constant that characterizes the
	rate of expansion in our Universe,
	$\ave{u_t} = \gamma_t \dot R_G = \gamma_t H$~\cite{comment}. 
	This emphasizes the similarity between the Big Bang of our Universe  
	and the Little Bangs of heavy ion collisions.

	The invariant single particle spectrum is obtained~\cite{3d,cs-review}
	from BL-H as
\bea
        N_1({\bf k}) & = &
                {\dst d^2 n\ov  2 \pi m_t dm_t\, dy  } \,  = \, 
                {\dst g \ov (2 \pi)^3} \, \overline{E} \, \overline{V} \,
        \overline{C}
        \, {
        1 \ov
        \exp\l({\dst  u^{\mu}(\overline{x})k_{\mu}
        \ov  T(\overline{x})} -
        {\dst \mu(\overline{x}) \ov  T(\overline{x})}\r) + s},
        \label{e:bl-n1-h}
\eea                                                         
	where all the terms have an intuitive, but mathematically well
	defined meaning,
	see ref.~\cite{cs-review} for more detailed definitions.                  
	The invariant form of  the two-particle 
	Bose-Einstein correlation function (HBT) of the BL-H
	is found in the binary source formalism ~\cite{3dcf98,cs-review} as:
\bea
        C_2({\bf k}_1,{\bf k}_2)
        & = &
        1 +
         \lambda_* \, \Omega(Q_{\parallel}) \,\,\,
                \exp\left(- Q_{\parallel}^2 \overline{R}_{\parallel}^2
                 - Q_{=}^2 \overline{R}_{=}^2
                 - Q_{\perp}^2 \overline{R}_{\perp}^2 \right).
\eea
	The dependence of the fit parameters on the 
	value of the mean momentum of the pair is suppressed,
	see ref.~\cite{cs-review} for the complete set of definitions
	of the variables and the radius parameters
	in terms of the BL-H model parameters.  
	The pre-factor $\Omega(Q_\parallel)$ of the 
	BECF induces oscillations within the Gaussian envelope 
	as a function of $Q_\parallel$. This oscillating
	pre-factor satisfies $0 \le \Omega(Q_\parallel) \le 1$ 
	and $\Omega(0) = 1$.  
	In practice, the period of oscillations is larger,
	 than the corresponding Gaussian radius, so the oscillations 
	are difficult to resolve.
	The above BL-H form of the two-particle correlation is
	explicitly boost-invariant, as all the relative momentum
	dependent variables $Q_\parallel$, $Q_=$, $Q_\perp$
	and all the corresponding radius parameters,
	$\overline{R}_\parallel$, $\overline{R}_=$, 
	$\overline{R}_\perp$ are defined 
	in an explicitely boost invariant manner. 
	The BL correlation function can be 
	equivalently expressed in the frequently used, 
	but not invariant Bertsch-Pratt (BP) form in the 
	LCMS frame~\cite{lcms}, 
	within the $\Omega=1$ approximation: 
\bea
	    C_2({\bf k}_1,{\bf k}_2)   & = &
        1 +  \lambda_* \exp\left[ - R_{\rm s}^2 Q_{\rm s}^2 - 
	R_{\rm o}^2 Q_{\rm o}^2
        - R_{\rm l}^2 Q_{\rm l}^2 - 
	2 R^2_{\rm ol} Q_{\rm o} Q_{\rm l} \right]. 
\eea
        The above formulas for the BECF and IMD, as were used in the
        fits, have been introduced in
        refs.~\cite{3d,3dqm,ster-cf98,na22}, and summarized recently
	in ref.~\cite{cs-review}. The analytic 
	formulas that relate the BL-H model parameters
	to the above forms for the spectra and correlation functions,
	are given by eqs.~(84-105), (115-118) and (129-140) 
	of ref.~\cite{cs-review}. Note, however, that eq. (132) of
	ref.~\cite{cs-review} contains an unfortunate misprint,
	$(\langle u_t\rangle + \langle \Delta T/T\rangle_r)$
	in the denominator should be replaced by 
	$(\langle u_t\rangle^2 + \langle \Delta T/T\rangle_r)$,
	so the correct form of eq. (132) of ref.~\cite{cs-review} 
	given explicitely by eq.~ (\ref{e:rxbar}) of the present manuscript.

\section{Buda-Lund fits to RHIC-1 spectra and correlations
	\label{s:3}}

        Here, we reconstruct the space-time picture
        of particle emission in Au + Au collisions at RHIC
	within the BL-H framework, by fitting {\it simultaneously}
	the PHENIX and STAR final data
        on two-particle correlations and single-particle spectra
	presented in refs.~\cite{phenix_imd,phenix_hbt,star_hbt}.
	The BL model, in certain domain of the 
	parameter space~\cite{3d}, features a scaling limiting 
	behavior of all the HBT radius parameters,
	$R_{\rm out} \simeq R_{\rm side} \simeq R_{\rm long} \approx 
	\tau_0\sqrt{\frac{T_0}{m_t}}$. 
	A unique feature of the BL model is that it 
	has been successfully tested against a detailed description 
	of the single-particle spectra and the two-pion
	Bose-Einstein correlation functions in {\it both} $h + p$ and $Pb+Pb$
	collisions at CERN SPS bombarding energies~\cite{na22,ster-qm99}.

	The measured and the calculated single particle spectra
	are connected with the help of a core-halo correction factor
	$\propto$ ${1 / \sqrt{\lambda_*}}$ , where 
	the experimental values of the intercept parameter
	$\lambda_* (y,m_t)$ are to be taken from 
	the measurements.
	In the lack of these $\lambda_*(y,m_t)$  values
	we have utilized their average $\lambda_*$ for 
	a core-halo correction when fitting the PHENIX $K^-$ and
	$p^-$ spectra. 
	In particular, the following average values were used  
	for the various particle types:
	$\overline \lambda_*(K) = 0.80 $ (estimated from the
	NA44 data on kaon-kaon correlations at CERN SPS~\cite{na44-kk-prl},
	$\overline \lambda_*(\overline{p}) = 0.995$ (the fraction
	of long lived resonances that decay to anti-protons is neglected).
	For pions, we have utilized the $\lambda_*(m_t)$ values  given in 
	refs.~\cite{star_hbt,phenix_hbt}.

\subsection{Improving on earlier results}

	Note also that we have performed
	the data analysis within a Gaussian approximation to the
	oscillating prefactor, improving on our earlier 
	results~\cite{bl-datong} ,
	where we have utilized the $\Omega =1$ approximation.
	As the oscillating prefactor $\Omega$ depends only on $Q_\parallel$,
	a Gaussian approximation to $\Omega$ results in a
	a Gaussian form  of the Bose-Einstein correlation
	function in the Buda-Lund variables, but with a 
	modification of the invariant longitudinal size of the
	source,
\be 
	\overline{R}_{\parallel,\Omega}^2 = 
		\overline{R}_\parallel^2 \left[1 + 
		\frac{\Delta\overline{\eta}^2}{\cosh^2(\overline{\eta})}\right],
\ee
	where $\overline{R}_\parallel^2$  is given by eq.~(138) of
	ref.~\cite{cs-review}. This modified invariant longitudinal
	radius parameter, $
	\overline{R}_{\parallel,\Omega}^2$ replaces
	$\overline{R}_{\parallel}^2$ in eqs. (116-118), when making
	the transformation from the Buda-Lund radius parameters	
	to the experimentally determined Bertsch-Pratt radius
	parameters following the lines of ref.~\cite{cs-review}.

	Furthermore, we improved on our
	earlier results~\cite{bl-datong} 
	by taking into account an
	$m_t$ dependent core-halo correction for the PHENIX spectra
	and correlations, and by fitting the absolute normalization
	of the single particle spectra in both experiments, properly
	utilizing the degeneracy, fugacity and  quantum statistical factors.
	This allows us to extract the chemical potential in the
	center of the fireball, in contrast to our 
	earlier fits~\cite{bl-datong}
	where the absolute normalization of the particle 
	spectra and the central value of the chemical potential
	distribution were not yet determined. In ref.~\cite{ster-jhep},
	we have attempted to determine the central value of the chemical
	potentials from the preliminary data. We improve also on this
	analysis by using the final, published data, 
	by releasing the central value of the pion
	chemical potential, that was previously fixed to 0,
	and by using the correct value of the $g = 2 s + 1 $ 
	the degeneracy factors, i.e.
	$g = (1,1,2) $ for ($\pi^-$, $K^-$, $p^-$).

\begin{figure}[htb]
\vspace{13.cm}
\begin{center}
\includegraphics{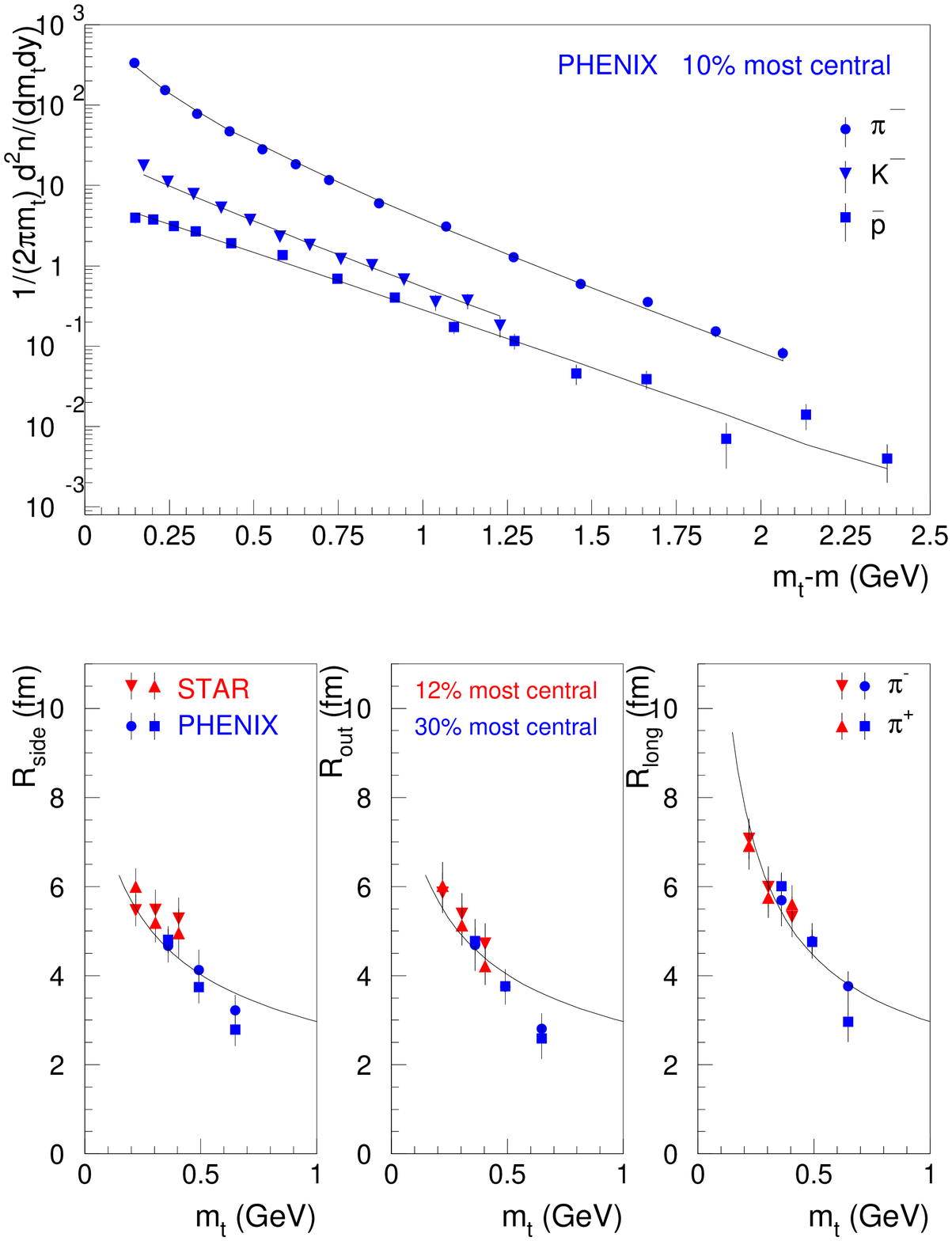}
\end{center}
\vspace{-1.5cm}
\caption[]{\label{figrhic-1}
        Simultaneous fits to the final PHENIX particle spectra and
	the final  PHENIX and STAR HBT radius parameters within the
	framework of the Buda-Lund hydro model. Solid line stands
	for the best Buda-Lund fits using parameter set Fit-I,
	as given in the first column of Table 1. The quality of the fit
	is statistically acceptable, CL = 29 \%.
	}
\end{figure}

\begin{figure}[htb]
\vspace{13.cm}
\begin{center}
\includegraphics{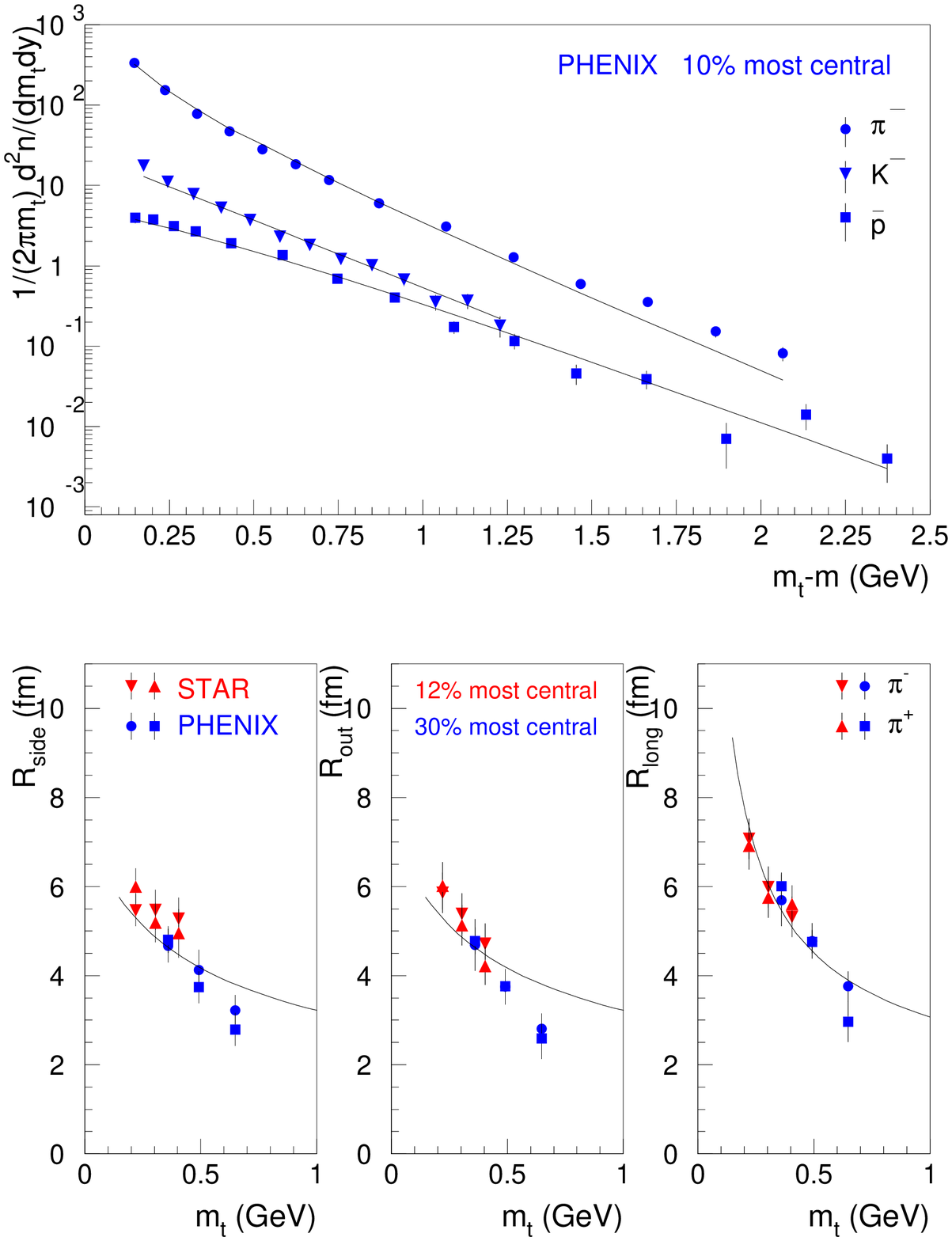}
\end{center}
\vspace{-1.5cm}
\caption[]{\label{figrhic-2}
        Simultaneous fits to the final PHENIX particle spectra and
	the final  PHENIX and STAR HBT radius parameters within the
	framework of the Buda-Lund hydro model, with $T_0=110$ MeV fixed. 
	Solid line stands
	for the Buda-Lund fits using parameter set Fit-III,
	as given in the third column of  Table 1. Although the fit
	does not look too bad, statistically the fit is not acceptable,
	as the confidence level of the fit is as low as $10^{-5}$.
	When comparing with Fig. 1, it is clear that the $m_t$ dependences
	of the side and the out radius parameters are not strong enough
	and the number of pions with high transverse momenta is too small.
	}
\end{figure}

	Fig. 1 illustrates the best combined  fit to the
	single particle spectra of negative pions, kaons and anti-protons of
	PHENIX,
	as well as  to the  transverse mass dependent HBT radii of STAR and
	PHENIX.
        Thus the hypothesis, that
        pions, kaons and protons are emitted from the same hydrodynamical
        source is in a good agreement with the fitted data, and
	the PHENIX and the STAR datasets are compatible with one another. 
	The parameters of the combined fit to PHENIX and STAR data
	are summarized in Table 1.

	A very important improvement over our earlier results
	is that now we have utilized the final PHENIX and STAR data points,
	in contrast to the earlier results, where the preliminary data points
	were utilized. 
	Unique minima are found and a statistically acceptable $\chi^2$/NDF 
	is obtained for the combined fit of both PHENIX and STAR datasets.
	Furthermore, the $\chi^2/NDF$ values are good not only on the
	combined data set, but on each of the fitted spectra and HBT radii,
	perhaps with the exception of the out radius parameter $R_{\rm o}$
	at the largest tranverse mass value, where the fit overestimates
	the measured point by about 2.5 standard deviations.

	The improvements did lead to a refinement of the 
	fitted parameters and their values, but, typically, the changes
	were within 3 standard deviations of the errors, with the
	exception of the central value of the freeze-out temperature,
	which increased significantly, from the preliminary
	RHIC average of $T_0 = 142 \pm 4$ MeV to
	the final  $T_0 = 202 \pm 13$ MeV, which is 
	a significant, close to 5 standard deviation modification. 
	Due to the above improvements, 
	the analysis of the final STAR and PHENIX data 
	now indicates the existence
	of a hot central part of the core, which was not resolved in
	the preliminary analysis. Clearly, more data points 
	would be very useful to confirm or invalidate this observation.
	In particular, HBT radius parameters of pions and kaons in the
	$m_t \approx 1$ GeV domain would provide very useful 
	constraints on the parameters of the BL-hydro model.

\begin{table}
\begin{tabular}{|l|rl|rl|rl|}
\hline
                 BL hydro \hspace{1.0cm}
                 & \multicolumn{2}{c|}{STAR+PHENIX}
                 & \multicolumn{2}{c|}{STAR+PHENIX}
                 & \multicolumn{2}{c|}{STAR+PHENIX} \\
\cline{2-7}
                 parameters
                 & \multicolumn{2}{c|}{I.}
                 & \multicolumn{2}{c|}{II.}
                 & \multicolumn{2}{c|}{III.} \\
\hline
$T_0$ [MeV]           & \hspace{0.3cm} 202    &$\pm$ 13
                      & \hspace{0.3cm} 140    & fixed 
                      & \hspace{0.3cm} 110    & fixed \\
$\langle u_t \rangle$ & 1.08   &$\pm$ 0.17
                      & 0.80   &$\pm$ 0.12
                      & 0.69   &$\pm$ 0.11 \\
$R_G$ [fm]            & 9.8    &$\pm$ 1.2
                      & 7.8    &$\pm$ 0.8  
                      & 7.3    &$\pm$ 0.8  \\
$\tau_0$ [fm/c]       & 6.1    &$\pm$ 0.3
                      & 7.6    &$\pm$ 0.2  
                      & 8.9    &$\pm$ 0.2  \\
$\Delta\tau$ [fm/c]   & 0.02   &$\pm$ 1.5
                      & 0.10   &$\pm$ 0.9 
                      & 0.07   &$\pm$ 0.6  \\
$\Delta\eta$          & 2.5    & fixed
                      & 2.5    & fixed     
                      & 2.5    & fixed     \\
$\langle {\Delta T 
\over T}\rangle_r$    & 0.84   &$\pm$ 0.24
                      & 0.10   &$\pm$ 0.04 
                      & 0.01   &$\pm$ 0.01 \\
$\langle {\Delta T 
\over T}\rangle_t$    & 1.3    &$\pm$ 0.4 
                      & 0.4    &$\pm$ 0.1 
                      & 0.1    &$\pm$ 0.1 \\
\hline
$\mu_0^{\pi^-}$
[MeV]                 & 75  & $\pm$ 19 
                      & 104 & $\pm$ 14  
                      & 118 & $\pm$ 12 \\ 
$\mu_0^{K^-}$ 
[MeV]                 & 107 & $\pm$ 14
                      & 172 & $\pm$ 17 
                      & 206 & $\pm$ 12 \\ 
$\mu_0^{\overline{p}}$
[MeV]                 & 305 & $\pm$ 41
                      & 412 & $\pm$ 17   
                      & 457 & $\pm$ 17 \\  
\hline
$\chi^2/$NDF          & \multicolumn{2}{l|}{74/68  = 1.09}
                      & \multicolumn{2}{l|}{99/69  = 1.43}
                      & \multicolumn{2}{l|}{127/69 = 1.84} \\
\hline
CL                    & \multicolumn{2}{c|}{28.9\%}
                      & \multicolumn{2}{c|}{1.0\%} 
                      & \multicolumn{2}{c|}{2.7e-3\%} \\
\hline

\end{tabular}
\caption
{
Source parameters from simultaneous fits of
final $Au+Au$ RHIC data of  PHENIX and STAR
on  particle spectra and HBT radius parameters with
the Buda-Lund hydrodynamical model. 
} 
\label{tab:results}
\end{table}

\begin{table}
\begin{center}
\begin{tabular}{|l|rl|rl|rl|}
\hline
                 BL hydro \hspace{1.0cm}
                 & \multicolumn{2}{c|}{Best fit}
                 & \multicolumn{2}{c|}{$T_0$ = 140 MeV}
                 & \multicolumn{2}{c|}{$T_0$ = 110 MeV}
                 \\
\cline{2-7}
                 parameters
                 & \multicolumn{2}{c|}{Surface values}
                 & \multicolumn{2}{c|}{Surface values}
                 & \multicolumn{2}{c|}{Surface values}
                 \\
\hline

$T_s$ [MeV]           & \hspace{0.3cm} 110 &$\pm$ 16 
                      & \hspace{0.3cm} 127 &$\pm$ 5 
                      & \hspace{0.3cm} 108 &$\pm$ 1 
	\\
$\langle \beta_t \rangle$ 
			& 0.73  &$\pm$ 0.06  
			& 0.62  &$\pm$ 0.04
			& 0.57  &$\pm$ 0.03
	\\
%\hline (This is valid at r_x=r_y = R_G)
%$\mu_s^{\pi^-}$        [MeV] 	& -126   & $\pm$ 23
%				& -36 	 & $\pm$ 15 
%				&   8  	&  $\pm$ 12 \\ 
%$\mu_s^{K^-}$          [MeV] 	
%				& -95 & $\pm$ 32 
%				&  33 & $\pm$ 17 
%				&  97 & $\pm$ 13 \\ 
%$\mu_s^{\overline{p}}$ [MeV] 
%				&  103& $\pm$ 51 
%				&  272& $\pm$ 17 
%				&  347& $\pm$ 17\\
\hline
%(This is valid at r_t=R_G)
$\mu_s^{\pi^-}$        [MeV] 	& -25	& $\pm$ 21
				&   34 	& $\pm$ 15 
				&   63 	& $\pm$ 12 \\ 
$\mu_s^{K^-}$          [MeV] 	
				&     6 & $\pm$ 29 
				&  103 & $\pm$ 16 
				&  152 & $\pm$ 12 \\ 
$\mu_s^{\overline{p}}$ [MeV] 
				&  204 & $\pm$ 49 
				&  343 & $\pm$ 17 
				&  402 & $\pm$ 17\\
\hline
Conf. level			
                 & \multicolumn{2}{c|}{28.9 \%}
                 & \multicolumn{2}{c|}{1.0 \%}
                 & \multicolumn{2}{c|}{2.7 $10^{-3}\%$}
                 \\
\hline

\end{tabular}
\end{center}
\caption
{
Calculated parameters, the surface values of the temperature,
flow and chemical potential distributions,
as evaluated from the simultaneous fits of Table 1 to
final $Au+Au$ RHIC data of  PHENIX and STAR
on  particle spectra and HBT radius parameters with
the Buda-Lund hydrodynamical model. The surface temperature  is
$T_s = T_0/(1 +\langle \Delta T/T\rangle_r)$, the
average 
surface three-velocity is 
$\langle \beta_t \rangle = \langle u_t\rangle /
	\sqrt{1 + \langle u\rangle_t^2}$ 
and the surface
chemical potentials are given by $\mu_s = \mu_0 - T_0/2$ for all
particles. 
} 
\label{tab:calc_results}
\end{table}

\section{Discussion \label{s:4}}
	We emphasize that the way how the data were presented
	are not ideal for fitting the BL-hydro model parameters.
	In particular, the BL-hydro model parameters can be  
	determined easier, if the rapidity dependent transverse
	mass spectra are given for as many particles, as possible,
	as the BL model contains terms that explicitely break
	the longitudinal boost invariance of the source.
	For an example of such an analysis, see ref.~\cite{na22}.
	The BL HBT radii also contain small correction terms
	that depend on the deviation from mid-rapidity 	
	in an explicite manner. The violation of complete
	longitudinal boost invariance with a finite space-time
	rapidity distribution is in fact  an important
	feature of the BL-hydro model, that can be utilized to 
	distinquish the BL-hydro from longitudinally boost invariant
	sources, like Bjoken's hydro solution, or, the blast-wave model
	of ref.~\cite{blastwave}. Note that the terminology
	``blast-wave" was introduced in ref.~\cite{s-blastwave},
	which generalized phenomenologically the Zim\'anyi-Bondorf-Garpman 
	(ZBG) solution~\cite{zbg} of non-relativistic hydrodynamics to
	relativistic flow profiles. It is very interesting to observe
	this as the BL-hydro model also corresponds to a relativistic
	extension of the ZBG solution.

	The major difference
	between the final state of heavy ion collisions
	at RHIC and at CERN SPS seems to be not only the 
	increased freeze-out time
	and the increased transverse flow or Hubble constant and
        relevantly larger tranverse geometrical radius at RHIC, but,
	most strikingly, the existence of a hot center located
	close to the beam axis, which evaporates particles with
	a temperature that is very close to, or above 
	the deconfinement temperature,
	$T_0 = 202 \pm 13$ MeV. This effect was not seen at CERN SPS,
	when the NA44, NA49 and WA98 data on particle spectra
	and correlations  were analyzed with the help of the BL-hydro 
	model~\cite{ster-qm99}. We have checked, that this minimum satisties
	eqs. (16-18) of ref.~\cite{na22}, i.e. the conditions for the
	validity of the saddle-point approximation are satisfied.

	We tried to determine the significance of this result by
	setting the central temperature to the value
	of $T_0 = 140$ MeV, artificially
	requiring that the temperature of the hottest zone at RHIC
	be similar to the value  of $T_0$, that were found in $h+p$ and 
	$Pb+Pb$ reactions at CERN SPS~\cite{na22,ster-qm99}. 
	 A reasonable fit was obtained, however,
	the confidence level of the fit decreased from 29 \%   to 1.2 \%,
	and the $\chi^2/NDF$ increased correspondingly, see the second column
	of Table 2.
	The blast-wave model describes the preliminary
	 transverse mass dependence of the
	STAR data on negative pions, kaons, anti-protons and anti-lambdas
	with $T_0 = 120 \null^{+50}_{-25}$ MeV, and $\langle \beta_t\rangle =
	0.52 \null^{+0.12}_{-0.08}$~\cite{snellings}. Within errors, 
	we recover this minimum from the combined fits to the PHENIX 
	final spectra and the final STAR and PHENIX HBT radius parameters,
	if we require a vanishing transverse temperature gradient in
	the BL-hydro model, or, if we fix by hand the central value of 
	the freeze-out temperature distribution to $T_0 = 110$ MeV,
	see the third column of Table 1. However, we find that
	this minimum is statistically not acceptable, as the confidence
	level of the fit is very small. 
	We have done another check, when determining the stability of the
	fit results for the variation of the central value of the freeze-out
	temperature. We have investigated the stability of the surface
	temperature for the variation of $T_0$ (best fit value versus
	$T_0 =140$ MeV fixed versus $T_0 = 110$ MeV fixed). 
	Interestingly, we found that the transverse
	temperature gradient, $\langle \Delta T/T\rangle_r$ 
	is strongly correlated to $T_0$,
	so a higher central temperature in the fit implies
	a larger temperature gradient, in such a way, that the
	surface temperature, $T_s = T_0/(1 + \langle \Delta T/T\rangle_r)$ 
	remains within errors the same. We have also calculated the
	3-velocity of the matter on the surface in the transverse
	direction, 
	$\langle \beta_t \rangle = \langle u_t\rangle /
	\sqrt{1 + \langle u\rangle_t^2}$ for a comparison,
	and the chemical potentials on the $r_t=R_G$ surface,
	that  are given by $\mu_s = \mu_0 - T_0/2$ for all particles.
	The freeze-out temperature and the average transverse
	velocity at $r_t=R_G$ are remarkably stable parameters of the fits,
	but, only the results in the first column of Table 2
	are statistically acceptable. According to this column, the 
	chemical potential of negative pions and kaons approximately
	vanishes at the $r_t = R_G$ surface, and the chemical potential
	for anti-protons is significantly bigger, than zero on the surface. 

	Furthermore, the surface temperature and the surface three-velocity
	was found to be similar to the average freeze-out temperatures
	that were obtained from the Regensburg model in ref.~\cite{boris}
	and the blast-wave model analysis of the RHIC final state in
	ref.~\cite{snellings}.	Our results are qualitatively as
	well as quantitatively similar to the findings of Florkowski
	and Broniowski~\cite{fb}, 
	who analyzed the PHENIX and STAR single particle
	spectra at RHIC in a hydro model that includes a spherically
	symmetric Hubble flow and similar in spirit to the BL-hydro 
	and the blast-wave models. They find that a value of
	$T_0=165 \pm 7$ MeV, $\mu_b = 41 \pm 5$ MeV and 
 	an average transverse flow velocity of
	$\langle \beta_t\rangle = 0.49$ describes the STAR and PHENIX
	preliminary data on the single particle spectra not
	only for negative pions, kaons and anti-protons, but
	also for $\phi$ mesons and $\Lambda$-s, 
	$\overline{\Lambda}$-s, and $K^*$-s,~\cite{fb-2}.

	We have found a non-vanishing chemical potential for  negative
	pions, kaons and anti-protons
	in the center of the fireball from the absolutely normalized
	single-particle spectra. The pion and kaon data were well described
	in all cases with a chemical potential that (within errors) 
	vanishes on the $r_t = R_G$ surface of the fireball. 
	These values together with
	the inhomogeneous chemical potential distribution
	of eq.~(124) of ref.~\cite{cs-review} 
	indicate a clear deviation from
	chemical equilibrium in the hadronic
	final state, as reconstructed within the Buda-Lund hydro model. 

	The similarities and the differences between an
	effective Quark Matter (QM) stage and a Quark Gluon Plasma (QGP)
	phase have been summarized recently in 
	ref.~\cite{csorgo-qm}. 
	The observed short duration of particle emission and the large
	transverse flow at RHIC contradicts to the picture of a 	
	soft, long-lived, evaporative Quark Gluon Plasma phase,
	that would consist of massless quarks and gluons.
	However, the final state does not exclude 
	a transient, explosive, suddenly
	hadronizing Quark Matter phase, that could be characterized by 
	massive valence quarks, the lack of gluons as effective 
	degrees of freedom, and a hard equation of state.

\section{Buda-Lund Predictions - What Next? \label{s:5}}
	
	In the above fits, we have utilized only mid-rapidity
	data points as given by the PHENIX and STAR collaborations.
	Additional measurements provide more stringent
	restrictions on the value of the fit parameters.

\subsection{(Pseudo-)rapidity dependent measurements}
	Recently, BRAHMS has published the pseudorapidity
	distribution of charged particles~\cite{brahms_dnch}, 
	which helps 
	to restrict the value of the width parameter 
	$\Delta\eta$, and also to get a more precise
	handle on the difference between the central 
	and the surface temperature of the fireball,
	as the broadening of the pseudo-rapidity distribution
	depends on this parameter. Preliminary results indicate
	an agreement with the value of $\Delta\eta=2.5$
	utilized in the fits presented here.
	
	The Buda-Lund hydro model predicts 
	a specific a coupling between
	the rapidity and the transverse mass dependent 
	single particle spectra . In particular,
	due to the finite longitudinal size of the
	expanding fire-tube it predicts\cite{3d,3dqm,cs-review}, 
	that
	the effective slope parameter decreases
	in the target and the projectile fragmentation
	regions as
\be
	T_{\rm eff}(y) = \frac{T_*}{1 + a (y - y_0)^2},
		\label{e:trap}
\ee
	where (transverse mass dependent)
	slope parameter at mid-rapidity is given by
\be
	T_* = T_0 + m_t \langle u_t\rangle^2 \frac{T_0 }
	{T_0 + m_t \langle \frac{\Delta T}{T}\rangle_r},
	\label{e:tstar}
\ee	
	and parameter $a$ is also expressed as a function of 
	the BL fit parameters in ref.~\cite{3d}. Note that
	in the original BL papers~\cite{3d,3dqm} the
	slope parameters were evaluated at $m_t =m$,
	in an approximation that neglected the transverse
	mass dependence of these values. If the fits 
	are done at low $p_t$, such an approximation
	can be warranted. If the temperature gradient
	effects are expected to be small, the phenomenological
	formula appears in its simplest form,
\be
	T_* = T_0 + m \langle u_t\rangle^2. \label{e:tsimple}
\ee	
	It would be important to experimentally test 
	the transverse mass dependence of the slope parameters,
	i.e. the difference between eqs. ~(\ref{e:tstar})
	and (\ref{e:tsimple}). It would also be very important
	to experimentally investigate the rapidity dependence
	of the single particle spectra, and the decrease
	of the slope parameters in the target and projectile
	fragmentation regions, as suggested by eq.~(\ref{e:trap}).

	Note that the BL hydro also predicts that the rapidity
	width of the double-differential invariant momentum
	distribution depends on the transverse mass in
	a specific manner, which is very sensitive to the
	central temperature $T_0$. In particular, the prediction is
\be
	\Delta y^2(m_t) = \Delta\eta^2 +\frac{T_0}{m_t}.
\ee
	Hence plotting the squared rapidity width of the single-particle
	momentum distribution as a function of $\frac{1}{m_t}$
	would provide a straight line, and its slope
	parameter would yield us an independent handle for the 
	value of the central temperature in the hotest and
	densest regions of the fireballs at RHIC. This information
	would be very valuable when concluding about the 
	temperature gradient effects and establishing the
	significance of a hot, central region with $T_0$ approximately
	the critical temperature of the quark-hadron phase
	transition, as suggested by the present analysis.

\subsection{Measurements at large $m_t$}
	More precise determination of the model parameters 
	will be possible if new data points become available
	for the effective source sizes (HBT radii) for pions
	and kaons at higher value of the transverse mass of the pair.
	The important point is that the $1/\sqrt{m_t}$ scaling
	of the HBT radius parameters is predicted to emerge
	in the high transverse mass limiting case, with small
	scaling violating terms, that play a larger role
	at small values of the transverse mass. Such a behavior
	is indeed seen in the BL fits in Fig. 1, in particular,
	the scaling violating terms are rather 
	apparent in the low transverse mass $R_l$ data points. 
	If the temperature gradient effects are not so important,
	the center of particle emission for particles
	with larger transverse masses moves more and more
	out from the $r_z$ axis, 
	hence more and more transverse
	flow effects result in a faster than $1/\sqrt{m_t}$ decrease
	in the out component. However, if the temperature
	gradient effects are important, all the 3 radius components
	will follow the $1/\sqrt{m_t}$ scaling law in the
	large transverse mass domain. A measurement of the 
	effective proton source as a function of $m_t$ and
	perhaps proton-deuteron coalescence measurements
	would be also tremendously useful to clarify the 
	status of this scaling law.

\subsection{Non-identical particle correlations}
	The BL model could be extended to study non-indentical
	particle correlations~\cite{ledniczky}, that provide
	very important keys to tell what kind of particles are 
	emitted first from the hot and dense decaying fireball.
	In particular, non-identical particle correlations
	are sensitive to both the temporal and spatial
	separations of the different kind of emitted particles.
	Preliminary data from the STAR collaboration indicates,
	that protons are emitted closer to the surface,
	than kaons, and kaons are emitted closer to the 
	surface, than pions. In the BL model, the production
	of heavier particles is more and more focussed
	to the collision axis, the transverse momentum
	and mass dependence of the center of particle production
	in the transverse plane is given by eq. (132) of ref.
	~\cite{cs-review}, which simplifies at mid-rapidity to
\be
	\overline{r_x} = R_G
		\frac{p_t \langle u_t\rangle}{T_0 + m_t
		(\langle u_t\rangle^2 + 
		\langle \frac{\Delta T	}{T}\rangle_r)},
		\label{e:rxbar}
\ee
	which implies that at any given value of the 
	transverse momentum $p_t$ the center of particle 
	production for heavier particles is closer to the
	collision axis, than that of the lighter ones.
	Note, that the BL-hydro also predicts,
	that the protons come
	from a smaller effective source than the pions,
	and due to this effect, some of the pions may
	appear from behind the protons.

\begin{figure}[htb]
\vspace{4.cm}
\begin{center}
\includegraphics{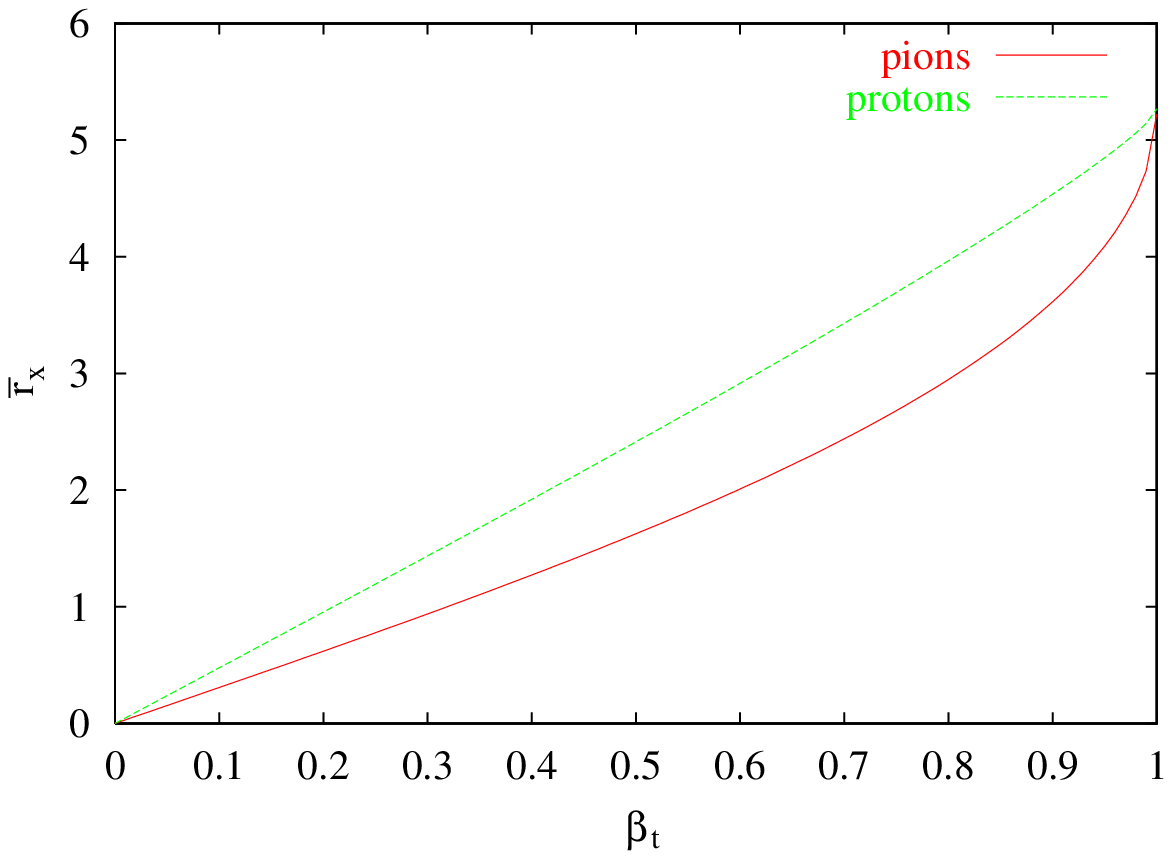}
\end{center}
\vspace{-1.5cm}
\caption[]{\label{f:nonid-1}
	The mean production point of pions and protons with
	a given velocity from the BL hydro model,
	corresponding to eq.~(\ref{e:rxbar}), utilizing the
	best fit values given in column I of Table 1. 
	Surprizingly, we find that
	within the BL-hydro model, heavier particles are emitted
	with larger transverse coordinate values, than lighter
	particles,  when non-identical particles with the same 
	velocity are compared. The BL hydro model predicts
	according to this figure, that both in the small and the large 
	transverse velocity limits the difference between the 
	transverse displacement of the pion and proton production points 
	disappears.
	}
\end{figure}

	In non-identical particle interferometry, 
	the effects are maximal if the velocity of
	the different particles are approximately the same. 
	It is striking to observe, that plotting 
	eq.~(\ref{e:rxbar}) as a function of the mean
	transverse velocity of the particle pair,
	heavier particles with a given velocity
	are emitted from a more forward region
	than the lighter particles! This feature of
	the BL model is qualitatively similar to the observations
	by the preliminary STAR data on non-identical
	particle correlation~\cite{szarvas,lisa}. 
 
	This feature of the BL model
	is closely related to the transverse
	mass scaling of the HBT radius parameters,
	and it would be very important to check experimentally.
	The BL model predicts a similar, focussing effect in
	the longitudinal direction too, namely that the
	center of emission of  heavier particles  at any
	given value of rapidity is closer to the midrapidity,
	than that of the lighter particles, 
\bea
	\overline{\eta} & = &\frac{y_0 - y}
		{ 1 + \Delta^2\eta \frac{m_t}{T_0}}, \\
	\overline{r}_z & = & \tau_0 \sinh(y+\overline{\eta}),\\
	\overline{t}_z & = & \tau_0 \cosh(y+\overline{\eta}),
\eea
	where $y$ is the rapidity of the particle,
	$y_0$ stands for the value of mid-rapidity
	and $\overline{\eta}$ is the space-time
	rapidity of the particle in the LCMS ($y=0$) frame,
	while the center of particle production is
	given in the frame of observation.
	Due to Lorentz boost effects, this result implies  
	that within the Buda-Lund picture,
	heavier particles are also emitted earlier,
	than the lighter ones, if the observation of the
	temporal and spatial sequence  is done
	in the LCMS ($y=0$) frame.

\section{Conclusions \label{s:6}}
        We find that the PHENIX and STAR data on single particle
        spectra of identified $\pi^-$, $K^-$ and $\overline{p}$ as well as 
        detailed $m_t$ dependent HBT radius parameters
        are consistent with the Buda-Lund hydro model as well as 
	with one another. 

	The major difference
	between the final state of heavy ion collisions
	at RHIC and at CERN SPS seems to be not only the 
	increased freeze-out time
	and the increased transverse flow or Hubble constant and
        relevantly larger tranverse gemetrical radius at RHIC, but,
	most strikingly, the existence of a hot center located
	close to the beam axis, which evaporates particles with
	a temperature that is very close to the deconfinement temperature,
	$T_0 = 202 \pm 13$ MeV. This effect was not seen at CERN SPS,
	when analysed in terms of the Buda-Lund hydro model.
	We find that this parameter can be determined independently from
	the temperature and the velocity of the surface, for which
	parameters we found rather conventional values.

	Note also that our findings are in qualitative agreement with
	numerical results found from Humanic's 
	cascade~\cite{humanic-sps,humanic-rhic} as well as
	from URQMD~\cite{bass}. 
	It seems that at RHIC-1, some of the pions, kaons
	and protons are emitted directly from a rather hot zone,
	that has (within the  errors of the reconstruction)
	the deconfinement temperature of QCD, 
	which even at finite chemical potential is
	predicted to be below the $T_c = 172 \pm 3$ MeV 
	value~\cite{fodor-katz}. 
	The hot center seems to be surrounded by a cooler 
	hadronic matter, with a surface temperature of
	about $T_s \approx 110$ MeV. Clearly, precision data from RHIC-2
	in a broad rapidity and transverse mass domain are
	needed to finalize the conclusions about the significance
	of this surprising result, in particular, the measurement
	of the pion and kaon HBT radius parameters at $m_t \approx 1 $ GeV
	would provide a very stringent restriction on the models.

\section*{Acknowledgment(s)}
	This research has been supported by the Hungarian OTKA 
	T038406, T034269, by a NATO Science Fellowship, and by the US NSF -
	Hungarian MTA-OTKA grant 0089462.

\section*{Notes}  
\begin{notes}
\item[a]
E-mails: csorgo@sunserv.kfki.hu, ster@rmki.kfki.hu
\end{notes}
%\vfill\eject


\begin{thebibliography}{99}
\bibitem{comment}
	Note that in ref.~\cite{ster-jhep} the $\gamma$ factor in
	this equation is not mentioned. It is important to emphasize,
	that $\langle u_t \rangle$ is an average {\it four-velocity}
	which includes a $\gamma $ factor: hence $\langle u_t\rangle$
	is not bounded from above.
\end{thebibliography}
\end{document}